\documentclass[12pt,preprint]{elsarticle}
\usepackage{graphicx}
\usepackage{epsfig}
\usepackage{amsmath}
\usepackage{amssymb}
\usepackage{color}
\usepackage{subfig}

\newcommand{\rmd}{\textrm{d}}

\usepackage{ulem} 

\newcommand{\Fermi}{\emph{Fermi-LAT}}

\biboptions{sort}
\journal{Astroparticle Physics}

\begin{document}

\begin{frontmatter}

\title{H.E.S.S. constraints on Dark Matter annihilations towards the Sculptor and Carina Dwarf Galaxies}
 \author{H.E.S.S. Collaboration}
 \author[ad1]{A.~Abramowski}
 \author[ad2]{F.~Acero}
 \author[ad3,ad4,ad5]{F. Aharonian}
 \author[ad6,ad5]{A.G.~Akhperjanian}
 \author[ad7]{G.~Anton}
 \author[ad8,ad9]{A.~Barnacka}
 \author[ad10]{U.~Barres de Almeida\fnref{ulis}}
 \author[ad11]{A.R.~Bazer-Bachi}
 \author[ad12,ad13]{Y.~Becherini}
% ...

\cortext[ach]{Corresponding authors}
\address[ad1]{
Universit\"at Hamburg, Institut f\"ur Experimentalphysik, Luruper Chaussee 149, D 22761 Hamburg, Germany}
\address[ad2]{
Laboratoire de Physique Th\'eorique et Astroparticules, Universit\'e Montpellier 2, CNRS/IN2P3, CC 70, Place Eug\`ene Bataillon, F-34095 Montpellier Cedex 5, France}
\address[ad3]{
Max-Planck-Institut f\"ur Kernphysik, P.O. Box 103980, D 69029 Heidelberg, Germany}
\address[ad4]{
Dublin Institute for Advanced Studies, 31 Fitzwilliam Place, Dublin 2, Ireland }
\address[ad5]{
National Academy of Sciences of the Republic of Armenia, Yerevan }
\address[ad6]{
Yerevan Physics Institute, 2 Alikhanian Brothers St., 375036 Yerevan, Armenia}
\address[ad7]{
Universit\"at Erlangen-N\"urnberg, Physikalisches Institut, Erwin-Rommel-Str. 1, D 91058 Erlangen, Germany}
\address[ad8]{
Nicolaus Copernicus Astronomical Center, ul. Bartycka 18, 00-716 Warsaw, Poland }
\address[ad9]{
CEA Saclay, DSM/IRFU, F-91191 Gif-Sur-Yvette Cedex, France}
\address[ad10]{
University of Durham, Department of Physics, South Road, Durham DH1 3LE, U.K.}
\address[ad11]{
Centre d'Etude Spatiale des Rayonnements, CNRS/UPS, 9 av. du Colonel Roche, BP 4346, F-31029 Toulouse Cedex 4, France}
\address[ad12]{
Astroparticule et Cosmologie (APC), CNRS, Universit\'{e} Paris 7 Denis Diderot, 10, rue Alice Domon et L\'{e}onie Duquet, F-75205 Paris Cedex 13, France}
\address[ad13]{
Laboratoire Leprince-Ringuet, Ecole Polytechnique, CNRS/IN2P3, F-91128 Palaiseau, France}
% ...
\fntext[ulis]{supported by CAPES Foundation, Ministry of Education of Brazil}
\fntext[such]{supported by Erasmus Mundus, External Cooperation Window}

\begin{abstract}
The Sculptor and Carina Dwarf spheroidal galaxies were observed with the H.E.S.S. Cherenkov telescope array between January 2008 and December 2009. The data sets consist of a total of 11.8 and 14.8 hours of high quality data, respectively. No gamma-ray signal was detected at the nominal positions of these galaxies above 220 GeV and 320 GeV, respectively. Upper limits on the gamma-ray fluxes at 95\%~C.L. assuming two forms for the spectral energy distribution (a power law shape and one derived from dark matter annihilation) are obtained at the level of 10$^{-13}$ to 10$^{-12}$~cm$^{-2}$s$^{-1}$ in the TeV range. Constraints on the velocity weighted dark matter particle annihilation cross section for both Sculptor and Carina dwarf galaxies range from $\left\langle \sigma v \right\rangle\sim 10^{-21}$~cm$^3$s$^{-1}$ down to $\left\langle \sigma v \right\rangle\sim 10^{-22}$~cm$^3$s$^{-1}$ depending on the dark matter halo model used. Possible enhancements of the gamma-ray flux are studied: the Sommerfeld effect, which is found to exclude some dark matter particle masses, the internal Bremsstrahlung and clumps in the dark-matter halo distributions.
\end{abstract}


\vspace{2pc}
\end{frontmatter}


\section{Introduction \label{sec:part1}}

The Dwarf Spheroidal Galaxies (dSphs) of the Local Group are the most common
satellites of the Milky Way and assumed to be gravitationally bound
dominantly by Dark Matter (DM). Although the predicted very high energy
(VHE, $E \gtrsim 100$~GeV) gamma ray flux from DM annihilation from dwarf
galaxies is smaller compared to the expected DM annihilation gamma ray
flux from denser regions of DM such as the Galactic Center, these
galaxies are promising targets for searches for gamma rays from DM
annihilation since they are environments with a favorably
low astrophysical gamma ray background. The galaxies themselves are
expected to contain no astrophysical gamma ray sources since no recent
star formation activity gives rise to VHE gamma rays (supernova
remnants, pulsar wind nebula, etc.) and little or no gas acting as
target material for cosmic rays has been measured~\cite{1998ARA&A..36..435M}. Additionally their position at high galactic latitude is well separated from the numerous sources of VHE gamma rays in the Galactic
plane. Also at such high altitudes  no detectable contamination due to diffuse VHE gamma ray emission is expected, which in any case has shown up so far only from the Galactic Center region~\cite{2006Natur.439..695A}.

The H.E.S.S. (\emph{High Energy Stereoscopic System}) array of Cherenkov telescopes has already observed dSphs and the collaboration has published results on the Sagittarius dSph~\cite{2008APh....29...55A,2010APh....33..274A} and the overdensity Canis Major~\cite{2009ApJ...691..175A}. In January 2008 H.E.S.S. launched observation campaigns on the Sculptor ($\alpha_{2000} = 01^{\rm h}02^{\rm m}19^{\rm s}_{^{\cdot}}2$, $\delta_{2000} = -33^{\circ}$~$33^{\prime}$~$00^{\prime \prime}_{^{\cdot}}0$, distance $\sim 79$ kpc,~\cite{1998ARA&A..36..435M}) and Carina ($\alpha_{2000} = 06^{\rm h}41^{\rm m}36^{\rm s}_{^{\cdot}}0$, $\delta_{2000} = -50^{\circ}$~$58^{\prime}$~$12^{\prime \prime}_{^{\cdot}}0$, distance $\sim 101$ kpc,~\cite{1998ARA&A..36..435M}) dSphs, which are among the most luminous dSphs near the Milky Way. The Sculptor dSph was discovered in 1938~\citep{1938BHarO.908....1S}, and was the first example of this type of galaxy in the vicinity of the Milky Way. The Carina dSph was discovered in 1977~\cite{1977MNRAS.180P..81C}. The best estimates of the orbits of the two dSphs show that Carina is likely to be more tidally disrupted than Sculptor~\cite{2003AJ....126.2346P,2006AJ....131.1445P}, leading to higher uncertainties for the DM content of the Carina dSph than of the Sculptor.

This paper presents the first results of a search for VHE gamma rays
from DM annihilation from the Sculptor and Carina dSphs. It is organized as follows: in Section~\ref{sec:part2} the analysis of the data is presented, from which upper limits on the gamma ray flux are extracted assuming power-law spectra  (Section~\ref{sec:part3}). In Section~\ref{sec:part4} flux limits for DM annihilation spectra are derived for both dSphs. Then the possibility of giving constraints on the DM particle properties is discussed (Section~\ref{sec:part5}), by assuming several DM galactic halo profiles of the dSphs, and considering various possibilities for the DM candidate particle, in particular those that could give rise to an enhancement to the gamma ray annihilation flux. The results obtained are discussed in Section~\ref{sec:part6}. The \Fermi\ (\emph{Fermi-Large Area Telescope}) collaboration~\cite{2009ApJ...697.1071A} has recently published a search for gamma ray emission from the Sculptor dSph in the energy range from 100~MeV to 100~GeV, the results from \Fermi~\cite{2010ApJ...712..147A} will also be addressed in Section~\ref{sec:part5}.

\section{H.E.S.S. observations and analysis \label{sec:part2}}

\subsection{The H.E.S.S. instrument}

The H.E.S.S. experiment is an array of four identical imaging atmospheric Cherenkov telescopes, observing VHE gamma rays. H.E.S.S. is located in the Khomas Highland of Namibia ($23^{\circ} 16^{\prime} 18^{\prime \prime}$ South, $16^{\circ}30^{\prime}00^{\prime \prime}$ East) at an altitude of 1800~m above sea level. Each telescope has an optical reflector consisting of 382 round facets of 60~cm diameter each, yielding a total mirror area of 107~m$^2$~\cite{2003APh....20..111B}. The Cherenkov light, emitted by charged particles in the electromagnetic showers initiated by primary gamma rays, is focused on cameras equipped with 960 photomultiplier tubes, each one subtending a field-of-view of $0.16^\circ$~\cite{2003ICRC....5.2887V}. The large field-of-view ($\sim$$5^\circ$) permits survey coverage in a single pointing. The direction and the energy of the primary gamma rays are reconstructed by the stereoscopic technique~\cite{2004APh....22..285F}.

\subsection{Observations and data analysis \label{parag:analyse}}

The observations of the Sculptor and Carina dSphs were conducted between January 2008 and December 2009. They were performed in \emph{wobble mode}~\cite{2006A&A...457..899A}, i.e. with the target typically offset by $0.7^\circ$ to $1.1^\circ$ from the pointing direction, allowing simultaneous background estimation in the same field-of-view. The data used for the analysis were taken at average zenith angles of $\sim$$14^\circ$ and $\sim$$34^\circ$ for the Sculptor and Carina dSphs, respectively, leading to different effective energy thresholds. A minimal energy ($E_{\rm min}$) is defined as the energy for which the acceptance of the instrument reaches 20\% of its maximum value, which for the Sculptor and Carina gives $E_{\rm min}\sim 220$~GeV and $E_{\rm min}\sim 320$~GeV, respectively. The data sets used for the analysis include only the observations that meet standard quality control criteria~\cite{2006A&A...457..899A}, based on the weather and data acquisition conditions.  The total data set amounts to $11.8$~h for Sculptor and $14.8$~h for Carina of live time after the quality selection. These parameters are summarized in Table~\ref{tab:data}.

\begin{table}[h]
  \begin{center}
    \begin{tabular}{|l|p{1.5cm}|p{1.5cm}|c|}
      \hline
      dSph Galaxy  & \multicolumn{2}{c|}{\bf Sculptor} & {\bf Carina}    \\
      \hline
      Observation Period   & \multicolumn{2}{c|}{2008 Oct - 2008 Nov} & 2008 Jan - 2009 Dec \\
      Live time (h) & \multicolumn{2}{c|}{11.8} &  14.8  \\
      \hline
      N$_{\rm ON}$ & \multicolumn{2}{c|}{117} &  86 \\
      \hline
      N$_{\rm OFF}$ & \multicolumn{2}{c|}{2283} & 1858 \\
      \hline
      $\alpha$ & \multicolumn{2}{c|}{0.04} &  0.05 \\
      \hline
      Significance & \multicolumn{2}{c|}{$1.0\, \sigma$} &  $-1.4\, \sigma$ \\
      \hline
      $N_{\gamma,\, {\rm tot}}^{95\%\,{\rm C.L.}}$ & \multicolumn{2}{c|}{32.4} &  8.6 \\
       \hline
      $E_{\rm min}$ (GeV)   &  \multicolumn{2}{c|}{220} & 320 \\
      \hline
      $N_{\gamma}^{95\%\,{\rm C.L.}}(E_{\gamma} > E_{\rm min})$ & \multicolumn{2}{c|}{19.2} & 10.2 \\
      \hline
    \end{tabular}
  \end{center}
  \caption{H.E.S.S. observation characteristics and upper limits on the observed number of gamma rays for the Sculptor and Carina dSphs. N$_{\rm ON}$ and N$_{\rm OFF}$ are the number of gamma-ray candidate events in the signal region and in the background region, respectively. $\alpha$ is defined as the ratio of the on-source area to the off-source area. The significance of the excess in the signal region is calculated for the given N$_{\rm ON}$, N$_{\rm OFF}$ and $\alpha$. $N_{\gamma,\, {\rm tot}}^{95\%\,{\rm C.L.}}$ is the 95\% confidence level upper limits on the total observed numbers of gamma-rays, and $N_{\gamma}^{95\%\,{\rm C.L.}}(E_{\gamma} > E_{\rm min})$ is the 95\% confidence level upper limits on the observed numbers of gamma-rays above the given minimal energy $E_{\rm min}$ for each dSph. The minimal energy $E_{\rm min}$ is defined as the energy for which the acceptance of the instrument reaches 20\% of its maximum value.    \label{tab:data}}
\end{table}

The data are analyzed using the \emph{model analysis} from~\cite{2009APh....32..231D}, with \emph{standard cuts}. This method is based on an accurate pixel per pixel comparison of the observed intensity with a pre-calculated semi-analytical model of showers. It provides an energy resolution of $\sim$$10 \%$ with an angular resolution at 2$\sigma$, defined as the 95\% containment radius, of the order of magnitude of $0.1^\circ$ in most of the energy range. The background was determined by the ring-background technique~\cite{2003APh....20..267P}, calculating the background for each position in the field-of-view using the background rate contained in a ring around the target.
\begin{figure}[t]
  \begin{center}
    \mbox{\hspace{0cm}\includegraphics[scale=0.3]{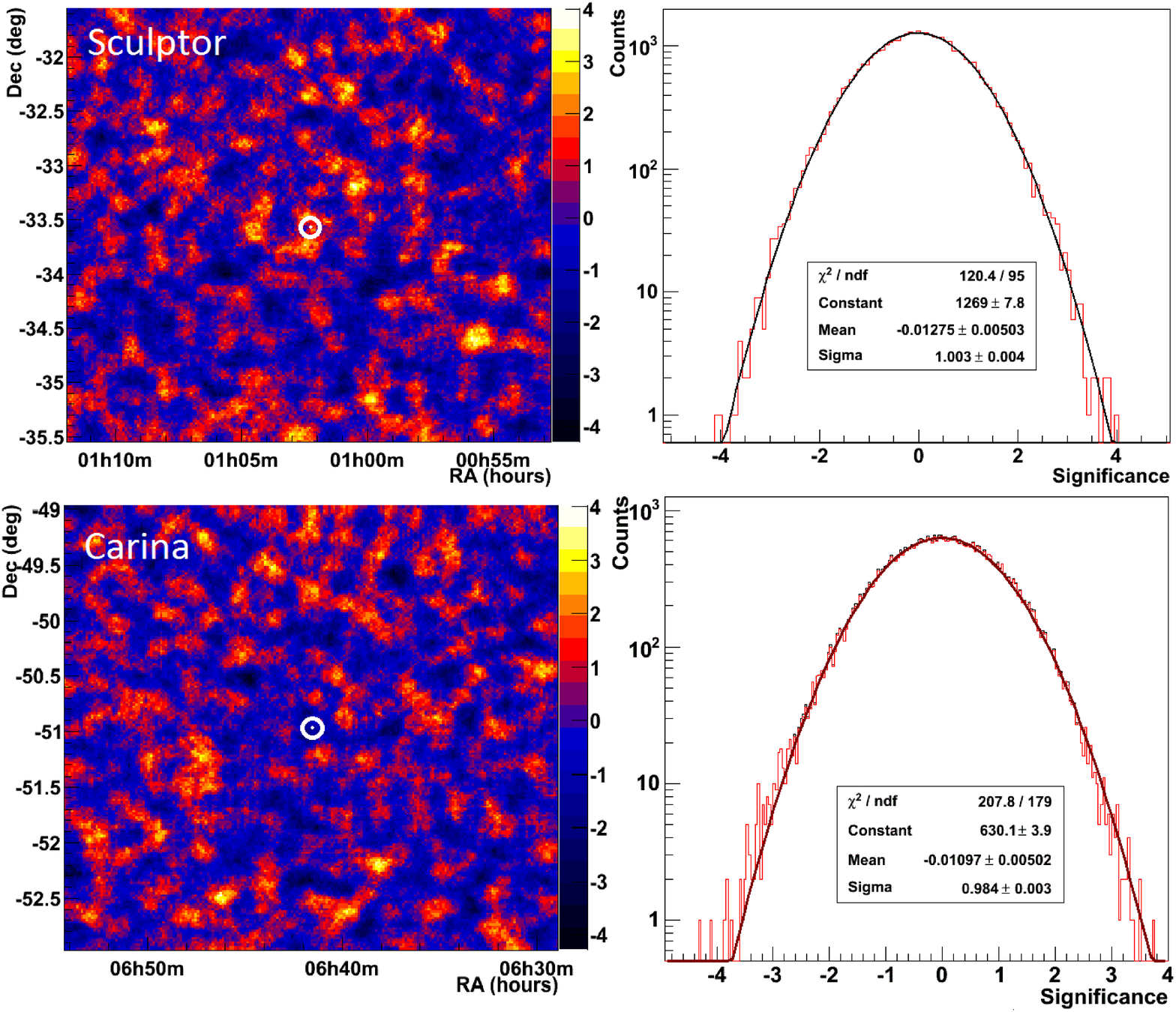}}
    \caption{Oversampled significance maps (with an integration region of $0.1^\circ$, white circle) in equatorial coordinates calculated according to the Li $\&$ Ma method~\cite{1983ApJ...272..317L} in the directions of Sculptor (top left) and Carina (bottom left) dSphs. Distributions of the significance in the maps for the Sculptor (top right) and the Carina (bottom right) dSphs. The solid line is a Gaussian fitted to the data. The significance distribution is well described by a normal Gaussian. No significant excess is seen at the target position.\label{fig:significance}}
  \end{center}
\end{figure}

No significant gamma ray excess was found above the estimated backgrounds at the nominal positions of Sculptor and Carina dSphs, as seen in Figure~\ref{fig:significance}. The significance of the excess in the $0.1^\circ$ radius integration area for Sculptor and Carina are $+1.0\, \sigma$ and $-1.4\, \sigma$, respectively\footnote{A sensitivity of 20.1 events is reported for Carina, following the suggestion in Feldman $\&$ Cousins~\cite{1998PhRvD..57.3873F} in the case where fewer events than the expected background are measured. The sensitivity is defined as the average upper limit (at 95$\%$ C.L.) obtained with the expected background and no true excess signal.}. This allows one to set 95\% confidence level upper limits on the total observed numbers of gamma-rays $N_{\gamma,\, {\rm tot}}^{95\%\,{\rm C.L.}}$ , using the number of gamma-ray candidate events in the signal region N$_{\rm ON}$, in the background region N$_{\rm OFF}$, and the ratio $\alpha$ of the on-source area to the off-source area. The limits have also been computed above the given minimal energy $E_{\rm min}$ for each dSph. The $N_{\gamma,\, {\rm tot}}^{95\%\,{\rm C.L.}}$ and $N_{\gamma}^{95\%\,{\rm C.L.}}(E_{\gamma} > E_{\rm min})$ obtained using the method described in Feldman $\&$ Cousins~\cite{1998PhRvD..57.3873F} are shown in Table~\ref{tab:data}.

\section{Flux upper limits with power-law gamma ray spectra \label{sec:part3}}

To obtain flux upper limits for standard astrophysical sources, power-law photon flux spectra of index $\Gamma$ are assumed,
\begin{equation}
  \label{eq:spectrum-pl}
  \frac{{\rm d}\Phi_{\gamma}}{{\rm d} E_{\gamma}} \propto E_{\gamma}^{-\Gamma}\, .
\end{equation}
The resulting $95\%$ confidence level upper limits on the gamma ray fluxes $\rm \Phi_{\gamma}^{95\%\,{\rm C.L.}}(E_{\gamma} > E_{\rm min})$ are derived through the knowledge of the acceptance of the instrument for each energy bin, given the parameters of the observation data sets, e.g. the mean zenith angle, the optical efficiency, the mean offset angle and the cuts used in the analysis~\cite{2006A&A...457..899A}. The index $\Gamma$ was varied between 1.8 and 2.4, which correspond to standard spectral indexes for astrophysical sources~\cite{2008RPPh...71i6901A}. The results are summarized in the Table~\ref{tab:phi95}.
\begin{table}[h]
  \begin{center}
    \begin{tabular}{|c|c|}
      \hline
      Galaxy & $\rm \Phi_{\gamma}^{95\%\,{\rm C.L.}}(E_{\gamma} > E_{\rm min})$ ($10^{-13}$~cm$^{-2}$s$^{-1}$) \\
      \hline
      Sculptor ($E_{\rm min} = 220$~GeV) & 5.1 - 6.2 \\
      Carina ($E_{\rm min} = 320$~GeV) & 1.6  - 2.0  \\
      \hline
    \end{tabular}
  \end{center}
  \caption{The $95\%$ confidence level upper limits on the gamma ray fluxes above the minimal energy $E_{\rm min}$, given in units of $10^{-13}$~cm$^{-2}$s$^{-1}$, for a power-law model with indices between $\Gamma = 1.8$ and $\Gamma=2.4$. The lower values of the upper limits are found for the index $\Gamma = 1.8$.  \label{tab:phi95}}
\end{table}

\section{Flux upper limits with Dark Matter annihilation spectra \label{sec:part4}}

\subsection{Theoretical framework}

Theories beyond the Standard Model (SM) of particle physics propose several candidates to account for the DM inferred from the gravitational interaction. Self-annihilation of DM particles into ordinary matter will in general produce gamma rays as sub-products, which may be detectable with current instruments.

For instance, some supersymmetric extensions of the SM~\cite{1996PhR...267..195J} predict the \emph{neutralino} as the lightest stable supersymetric particle, which is a good candidate for DM~\cite{1996PhR...267..195J}. Several parametrizations for the continuum contribution to the total neutralino self-annihilation gamma-spectrum are used here~\cite{1998APh.....9..137B,2004PhRvD..70j3529F}. The latter arises from the hadronisation and decay of the cascading products. Universal extra dimensional extensions of the SM provide also suitable DM candidates. In the Kaluza-Klein (KK) models the first KK mode of the hypercharge gauge boson $\widetilde{B}^{(1)}$ is a DM particle candidate~\cite{Servant:2002aq}.

The gamma ray flux from the annihilations of DM particles of mass $m_{\rm DM}$ in a DM halo is given by a particle physics term times an astrophysics term:
\begin{equation}
\label{eqnp}
\frac{\rmd\Phi_{\gamma}(\Delta\Omega,E_{\gamma})}{\rmd E_{\gamma}}\,=\frac{1}{8\pi}\,\underbrace{\frac{\langle
\sigma v\rangle}{m^2_{\rm DM}}\,\frac{\rmd N_{\gamma}}{\rmd E_{\gamma}}}_{\rm Particle\,
Physics}\,\times\,\underbrace{\bar{J}(\Delta\Omega)\Delta\Omega}_{\rm Astrophysics} \, ,
\end{equation}
where the astrophysical factor is defined as
\begin{equation}
\overline{J}(\Delta\Omega) = \frac{1}{\Delta \Omega} \int_{\Delta \Omega} \rmd\Omega \int_{\rm los} \rho^2[r(s)] \rmd s \, .
\label{jbar}
\end{equation}
In equation~\ref{jbar} the squared density of DM ($\rho^2$) is integrated along the line of sight (los) and over the solid angle $\Delta\Omega $. The solid angle is fixed as the angular resolution of the telescope for a point-like source search. For the H.E.S.S. experiment $\Delta \Omega = 10^{-5}$~sr. The particle physics term contains the DM particle mass, $m_{\rm DM}$, the velocity-weighted annihilation cross section, $\langle \sigma v\rangle$, and the differential gamma ray spectrum from all final states weighted by their corresponding branching ratios, $\rmd N_{\gamma}/\rmd E_{\gamma}$.

The number of detected gamma rays above a minimal energy $E_{\rm min}$ is given by
\begin{equation}
N_{\gamma} (E>E_{\rm min})= T_{\rm obs} \int_{E_{\rm min}}^{\infty} A_{\rm eff}(E_{\gamma})\, \frac{\rmd \Phi_{\gamma}}{\rmd E_{\gamma}} \, \rmd E_{\gamma} \, ,
\label{ngamma}
\end{equation}
where $T_{\rm obs}$ is the observation time, and $A_{\rm eff}(E_{\gamma})$ is the effective area of the detector as a function of the gamma ray energy, the zenith angle, the offset of the source from the pointing direction and the selection cuts. The differential flux $\rmd \Phi_{\gamma}/\rmd E_{\gamma}$, in the case of the DM annihilation, is given by equation~\ref{eqnp}.

\subsection{Flux upper limits with neutralino annihilation spectra \label{parag:indepedant}}

Before modelling the shape of the DM halos and addressing the astrophysics factor $\bar{J}(\Delta\Omega)\Delta\Omega$, upper limits on the flux at $95\%$~C.L. for different annihilation spectra are considered as a function of the DM particle mass. The number of gamma rays above a minimal energy can be related to the integral flux by
\begin{equation}
N_{\gamma}^{95\%\,{\rm C.L.}} (E>E_{\rm min}) = \frac{T_{\rm obs} \int_{E_{\rm min}}^{m_{\rm DM}} A_{\rm eff}(E_{\gamma})\, \frac{\rmd N_{\gamma}}{\rmd E_{\gamma}} \, \rmd E_{\gamma} }{\int_{E_{\rm min}}^{m_{\rm DM}} \frac{\rmd N_{\gamma}}{\rmd E_{\gamma}} \, \rmd E_{\gamma}} \times \Phi_\gamma^{95\%\,{\rm C.L.}} (E>E_{\rm min}) \, .
\label{eq:ngamma-sup-emin}
\end{equation}
The annihilation spectrum depends on the nature of the final states. A parametrization using the the average of the WW and ZZ final states was taken from Bergstr\"{o}m et al.~\cite{1998APh.....9..137B}, which will be used here and in Section~\ref{sec:part5}. For annihilation into ${\rm b}\overline{\rm b}$ and $\tau^+\tau^-$ PYTHIA 6.225~\cite{2001CoPhC.135..238S} was used to compute the spectra. The upper limit on the flux at $95\%$~C.L. is related to the upper limit on the number of observed gamma rays at $95\%$~C.L. by replacing the $N_{\gamma}^{95\%\,{\rm C.L.}}$ from Section~\ref{parag:analyse} (Table~\ref{tab:data}) in equation~\ref{eq:ngamma-sup-emin}.

\begin{figure}
  \centering
  {\label{fig:phi95-fermihess-dm}\includegraphics[scale=0.45]{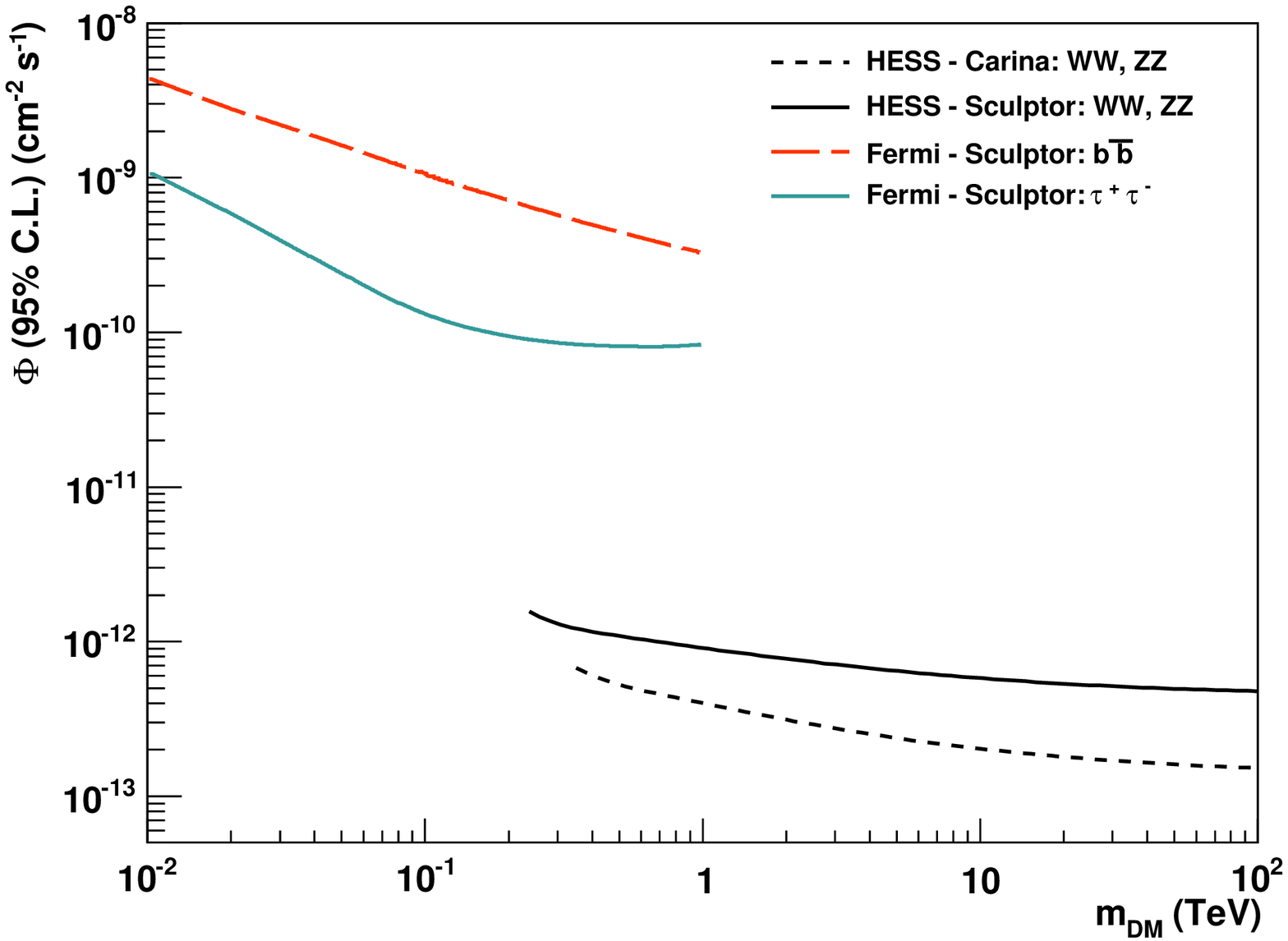}}\\
  {\label{fig:phi95-hess-dm}\includegraphics[scale=0.45]{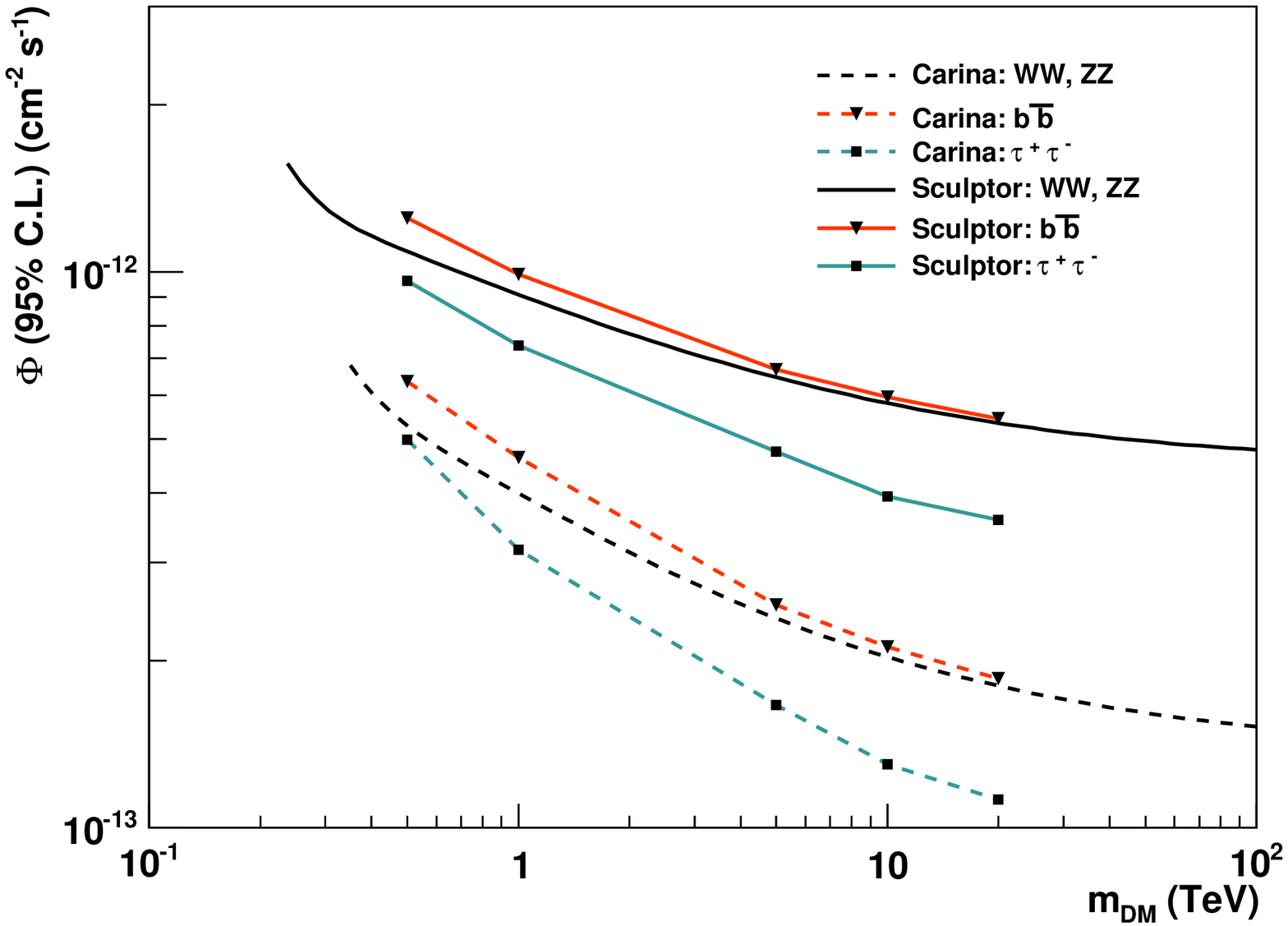}}
  \caption{\emph{(top)} Upper limits on fluxes at $95\%$~C.L. for the Carina (dashed lines) and Sculptor (solid lines) dSphs obtained with H.E.S.S., for $E>320$~GeV and  $E>220$~GeV, respectively, as function of the DM mass. The parametrization of the annihilation spectrum using the the average of the WW and ZZ final states is extracted from Bergstr\"{o}m et al.~\cite{1998APh.....9..137B}. \Fermi\ results for Sculptor~\cite{2010ApJ...712..147A} with $E \gtrsim 100$~MeV are also plotted for $\tau^+\tau^-$ and $b\overline{b}$ final states~\cite{2010ApJ...712..147A}. \emph{(bottom)} A zoomed view on the upper limit from H.E.S.S. using in addition various parametrizations for the annihilation spectrum~\cite{2001CoPhC.135..238S}.\label{fig:phi95-dm}}
\end{figure}

 Figure~\ref{fig:phi95-dm} shows the calculated upper limit on the flux for both Sculptor and Carina dSphs, which depends on the assumed spectrum and hence on the mass of the neutralino. The results obtained by \Fermi~\cite{2010ApJ...712..147A} for the Sculptor dSph and energies $\gtrsim 100$~MeV are also plotted. As can be seen, for high neutralino masses ($\gtrsim 500$~GeV) H.E.S.S. is more sensitive. The flux sensitivity is qualitatively driven by the product of the acceptance $A_{\rm eff}(E_{\gamma})$ times the observation time $T_{\rm obs}$. Using the acceptances of about $A_{\rm eff}\sim 10^5$~m$^2$ for H.E.S.S. and of a few m$^2$ for \Fermi, and observation times of about $\sim$$12$~hours for H.E.S.S. and $\sim$$11$~months for \Fermi, the ratio between their sensitivities for a given DM mass yields a better sensitivity for H.E.S.S. by a factor of a few hundred, for masses well above the H.E.S.S. threshold.

\section{Exclusion limits on the Dark Matter annihilation cross section \label{sec:part5}}

\subsection{Astrophysical factor calculation \label{parag:modelling}}

In order to calculate the exclusion limits on the DM annihilation cross section, one needs to model the density distribution of DM in the observed target that will be used in the astrophysical factor $\overline{J}$ calculation. Two hypotheses for spherical DM halo profiles are used for Sculptor and Carina: a pseudo-isothermal profile~\cite{Swaters:2000nt}, and the \emph{Navarro, Frenk, and White} (NFW) profile~\cite{1996ApJ...462..563N}.

The pseudo-isothermal sphere case belongs to the cored profile class. It is characterized by the mass density profile
\begin{equation}
\rho_{\rm ISO}(r)= \rho_{0}\frac{r_{\rm c}^{2}}{(r_{\rm c}^2+r^2)} \, ,
\end{equation}
where $r_{\rm c}$ is the core radius and $\rho_0$ is a normalization factor. This profile is completely defined by $\rho_0$ and $r_{\rm c}$.

The NFW profile is a cusped profile, with the mass density given by
\begin{equation}
\rho_{\rm NFW}(r) = \frac{\delta_{\rm c} \rho_{\rm c}^0}{(r/r_{\rm s})(1+r/r_{\rm s})^{2}} \, ,
\end{equation}
where $r_{\rm s}$ is a scale radius, $\rho_{\rm c}^0$ the present critical density of the Universe \footnote{$\rho_{\rm c}^0 = 2.775 \times 10^{11}~h^2~{\rm M}_{\odot} {\rm Mpc}^{-3} = 1.053 \times 10^{-5}~ h^2~({\rm GeV}/c^2)~{\rm cm}^{-3}$ \cite{pdg}} and $\delta_{\rm c}$ a characteristic over-density, defined by $\delta_{\rm c} = \Delta_{\rm v}\,c^3 g(c)/3$, with $\Delta_{\rm v}=200$. Here $c = R_{\rm v}/r_{\rm s}$ is the concentration parameter of the halo, where $R_{\rm v}$ is the virial radius, and $g(c) = (\ln (1+c) - c/(1+c))^{-1}$. In this case the profile is completely defined by the concentration parameter $c$ and the virial mass $M_{\rm v}$, expressed as
\begin{equation}
M_{\rm v}^{\rm NFW} = \frac{4\pi}{3}\Delta_{\rm v} \rho_{\rm c}^0 R_{\rm v}^3 \, .
\label{eq:mvir-nfw}
\end{equation}
Given a hypothesis about the gravitational potential of a galaxy, i.e. its DM halo profile, and about the velocity dispersion anisotropy of its stars, one can obtain the theoretical line-of-sight velocity dispersion from the Jeans equation. The methodology used to find each set of parameters for each halo profile used in the literature cited below consists of comparing the observed line-of-sight velocity dispersion of stars, $\sigma_{\rm los}$, for each distance bin with that predicted for the various models. Minimizing the $\chi^2$ between the model and the data provides the best set of parameters. $\sigma_{\rm los}$ is also dependent on the stars velocity dispersion anisotropy $\beta(r)$ (see e.g. equation~7 of~\cite{1982MNRAS.200..361B}). The latter is taken into account in the fitting procedure and different parametrizations were used.\\

{\bf Sculptor: }
The DM halo mass content of Sculptor dSph as well as its profile parameters are estimated in a recent paper~\cite{Battaglia:2008jz} and thesis~\cite{BattagliaThesis}, using two RGB (\textit{Red Giant Branch}) stars populations to partially break the Jeans degeneracy in the DM halo modelling~\cite{2009ApJ...701.1500A}. The last gravitationally bound star was measured at $r_{\rm last}=1.8$~kpc, which gives an estimate of the tidal radius~\cite{BattagliaThesis} and is used in the integration to obtain $\overline{J}(\Delta\Omega)$. Two hypotheses for the $\beta(r)$ profile were explored in~\cite{Battaglia:2008jz}: a radially constant velocity dispersion anisotropy, and a Osipkov-Merritt (OM) velocity dispersion anisotropy~\cite{1979PAZh....5...77O,1985AJ.....90.1027M}. Using the parameters extracted from~\cite{BattagliaThesis}, the astrophysical factor $\overline{J}$ is computed for eight different Dark Matter halos. The parameters as well as the astrophysical factor are summarized in the Table~\ref{tab:sculptor}. The large variety of investigated DM halos allows one to encompass the astrophysical uncertainties induced by the modelling.\\
\begin{table}[h!]
  \begin{center}
    \begin{tabular}{| l | c | c | c| c| }
      \hline
      \multicolumn{5}{|c|}{\textbf{Sculptor dSph}}\\
      \hline
      \hline
      Iso Profile & $r_{\rm c}$ & $M_{r_{\rm last}}$ & $\rho_0$  & $\overline{J}$  \\
              & (kpc)& ($10^{8}$~M$_{\odot}$)& ($10^{7}$~M$_{\odot}$kpc$^{-3}$) & ($10^{23}$~GeV$^2$cm$^{-5}$) \\
      \hline \hline
      $\beta=const$ & 0.05 & 1.2& 221 & 2.98 \\
      \cline{2-5}
      & 0.5 & 3.3 & 9.13  & 0.27 \\
      \hline
      $\beta=\beta_{OM}$ & 0.05 & 1.3 & 240  &  3.49 \\
      \cline{2-5}
      & 0.5 & 3.4 & 9.40  & 0.29\\
      \hline
      \hline
      NFW Profile & $c$ & $M_{\rm v}$ & $r_{\rm s}$  & $\overline{J}$  \\
      &  & ($10^{9}$~M$_{\odot}$)&  (kpc) & ($10^{23}$~GeV$^2$cm$^{-5}$)  \\
      \hline \hline
      $\beta=const$ & 20 & 1.9 & 1.26 & 2.75 \\
      \cline{2-5}
      & 35 & 0.59& 0.48 & 5.20 \\
      \hline
      $\beta=\beta_{\rm OM}$ & 20 & 2.2 & $1.32$ & 3.29 \\
      \cline{2-5}
      & 35 & 0.68 & 0.51 &  6.24 \\
      \hline
    \end{tabular}
    \caption{Structural parameters of the eight best fits~\cite{Battaglia:2008jz} in the case of a pseudo-isothermal and NFW DM halo profiles for the Sculptor dSph, as well as the corresponding values of the astrophysical factor $\overline{J}$, for two hypotheses of the velocity anisotropy profile $\beta(r)$, two core radii (pseudo-isothermal) and two concentration parameters (NFW). \label{tab:sculptor}}
  \end{center}
\end{table}

{\bf Carina: }
Here, the best fit parameters of each DM halo profile were given for a NFW profile in~\cite{2007ApJ...667L..53W} and for the pseudo-isothermal profile in~\cite{2007ApJ...663..948G}. The parameters for the NFW profile obtained from~\cite{2007ApJ...667L..53W} are the virial mass $M_{\rm v} = 2\times10^8$~M$_\odot$ and the star's velocity dispersion anisotropy $\beta(r) = -0.5$. Replacing $M_{\rm v}$ in equation~\ref{eq:mvir-nfw}, the virial radius is found to be $R_{\rm v} = 12$~kpc. Using the relationship between $M_{\rm v}$ and the concentration parameter $c$ found in~\cite{Jing:1999ir} (see also~\cite{2007ApJ...657..241K}) yields $c\simeq 22$. Finally $r_{\rm s}=5.35\times10^{-1}$~kpc is obtained using $r_{\rm s} = c/R_{\rm v}$. The parameters for a pseudo-isothermal profile with an isotropic velocity dispersion ($\beta(r) = 0$) and consistent with the kinematic data were obtained from~\cite{mark:com}, see also Figure~4 of~\cite{2007ApJ...663..948G}. The tidal radius is set arbitrarily to $r_{\rm t} = 2.0$~kpc, which is a conservative value \citep{2005AJ....130.2677M, 2006ApJ...649..201M}, and it is used in the integration to obtain $\overline J (\Delta\Omega)$. The parameters of the DM halo profiles as well as the astrophysical factor $\overline{J}$ are summarized in Table~\ref{tab:carina}.
\begin{table}[h!]
  \begin{center}
    \begin{tabular}{| l | c | c | c| c| }
      \hline
      \multicolumn{5}{|c|}{\textbf{Carina dSph}}\\
      \hline
      \hline
      Iso Profile & $r_{\rm c}$ & $r_{\rm t}$  & $\rho_0$  & $\overline{J}$  \\
        $\beta=0$   &(kpc)  & (kpc) & ($10^{8}$~M$_{\odot}$kpc$^{-3}$)  &($10^{22}$~GeV$^2$cm$^{-5}$) \\
      \cline{2-5}
      & 0.22 & 2.0 & 1.36 & 2.01 \\
      \hline
      \hline
      NFW Profile & $c$ & $M_{\rm v}$ & $r_{\rm s}$ & $\overline{J}$ \\
        $\beta=-0.5$  &  &  ($10^{9}$~M$_{\odot}$) &  (kpc) &  ($10^{22}$~GeV$^2$cm$^{-5}$) \\
      \cline{2-5}
      & 22 & 0.20 & 0.54 & 4.37 \\
      \hline
    \end{tabular}
    \caption{Structural parameters of the two best fits~\cite{2007ApJ...663..948G,2007ApJ...667L..53W} in the case of a pseudo-isothermal and NFW DM halo profile for the Carina dSph, as well as the corresponding value of the astrophysical factor $\overline{J}$ (see equation~\ref{jbar}).\label{tab:carina}}
  \end{center}
\end{table}

\subsection{Generic case for exclusion limits\label{parag:profile}}

\begin{figure}
  \centering
  \label{fig:exclusionSc}\includegraphics[scale=0.39]{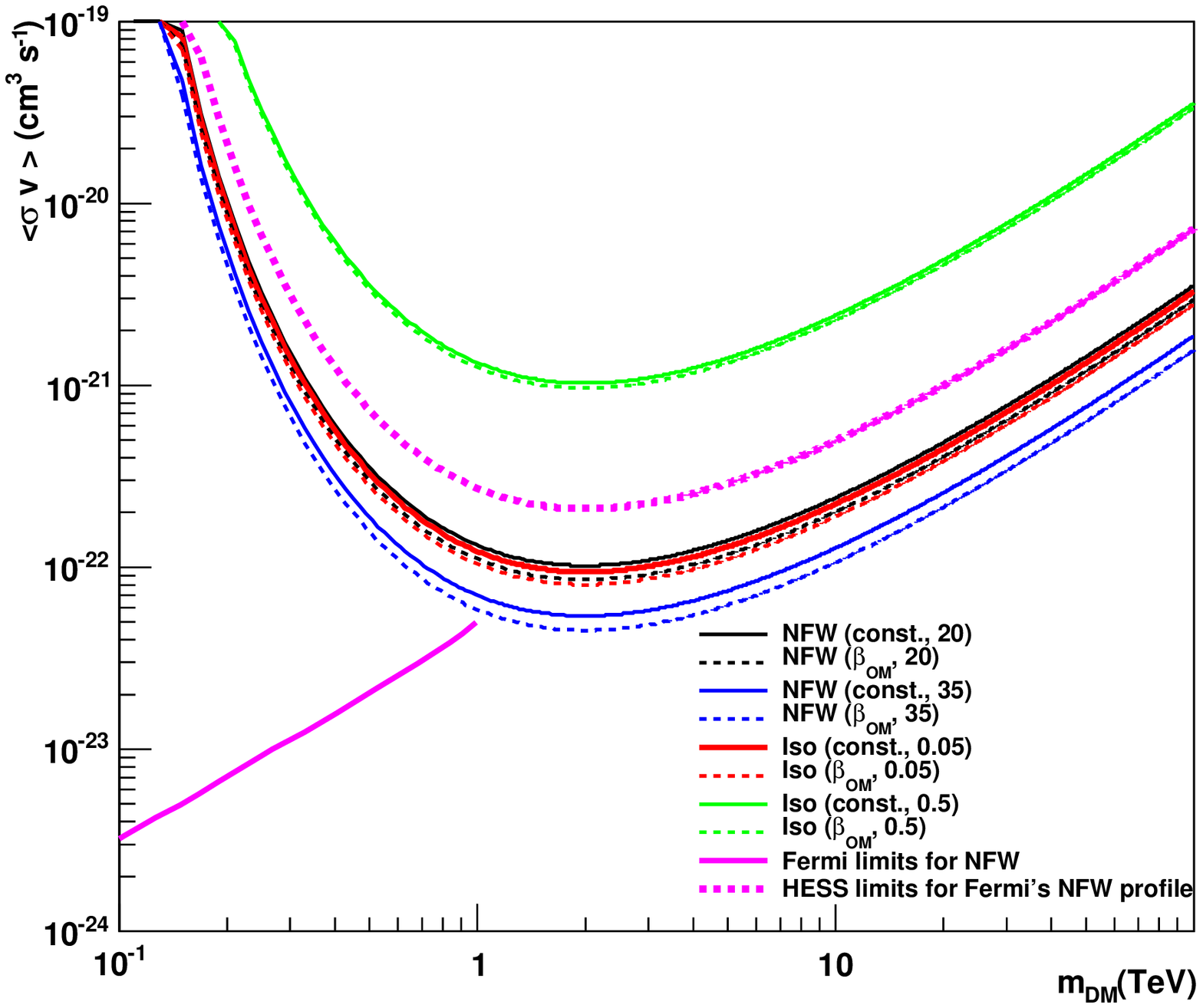}\\
  \label{fig:exclusionCa}\includegraphics[scale=0.39]{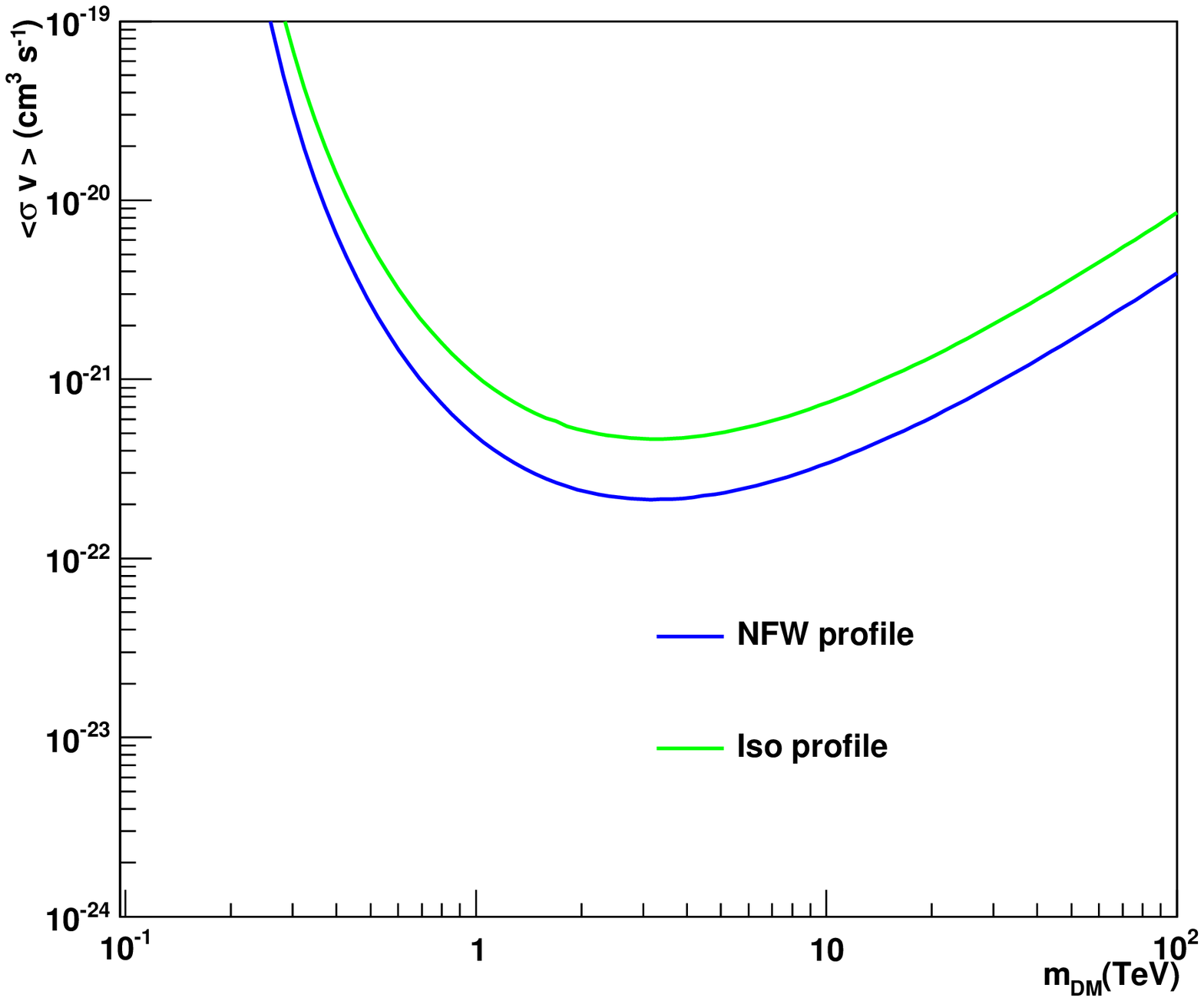}
  \caption{Upper limit at $95\%$ C.L. of $\left\langle \sigma v \right\rangle$ as function of the DM particle mass for different DM halos for Sculptor (\emph{top}) and Carina (\emph{bottom}). For the NFW halo profile of Sculptor two concentration parameters are used: 20 and 35. For the pseudo-isothermal halo profile two core radii are used: $0.05$~kpc and $0.5$~kpc. Two hypotheses on the velocity anisotropy parameter are also studied: a constant (solid lines) and an Osipkov-Merritt (dashed lines) anisotropy. The velocity anisotropy and the concentration parameters are given in brackets for the NFW profile. The velocity anisotropy and the core radius are given in brackets for the pseudo-isothermal profile. The \Fermi\ limits~\cite{2010ApJ...712..147A} for a NFW profile are also plotted as well as the H.E.S.S. limits for this NFW profile ($r_{\rm s}= 0.9$~kpc and $\rho_{\rm s} = 3.7\times 10^{7}$~M$_{\odot}$kpc$^{-3}$). For Carina both the NFW halo profile and the pseudo-isothermal halo profile are plotted (see text for parameters).\label{fig:exclusion}}
\end{figure}

The $95\%$ C.L. upper limit on the velocity-weighted annihilation cross section as function of the DM particle mass for a given halo profile is given by
\begin{equation}
\left\langle \sigma v \right\rangle_{\rm min}^{95\%\,{\rm C.L.}} = \frac{8\pi}{\overline{J}(\Delta \Omega)\Delta \Omega} \times \frac{m_{\rm DM}^{2}\, N_{\gamma,\, {\rm tot}}^{95\%\,{\rm C.L.}}}{T_{\rm obs}\, \int_{0}^{m_{\rm DM}} A_{\rm eff}(E_{\gamma}) \, \frac{\rmd N_{\gamma}}{\rmd E_{\gamma}}(E_\gamma) \, \rmd E_{\gamma}} \, ,
\end{equation}
where the parametrization of $\rmd N_{\gamma}$/$\rmd E_{\gamma}$ is taken from~\cite{1998APh.....9..137B} for typical neutralino self-annihilation into WW and ZZ pairs, and calculated from~\cite{Servant:2002aq} for Kaluza-Klein $\widetilde{B}^{(1)}$ self-annihilation. The exclusion curves for the neutralino case are plotted for the Sculptor and Carina dSphs in Figure~\ref{fig:exclusion} referring to the halo profiles given in the Table~\ref{tab:sculptor} for Sculptor, and in the Table~\ref{tab:carina} for Carina, respectively. The \Fermi\ exclusion limit for Sculptor is added extending up to 1~TeV~\cite{2010ApJ...712..147A}, which is based on a NFW profile with $r_{\rm s} = 0.9$~kpc and $\rho_{\rm s} = \delta_{\rm c} \times \rho_{\rm c}^0 = 3.7\times 10^{7}$~M$_{\odot}$kpc$^{-3}$, and a neutralino parametrization with only $b\overline{b}$ in the final state. Using the parameters $r_{\rm s}$ and $\rho_{\rm s}$ from \Fermi\ paper~\cite{2010ApJ...712..147A}, the astrophysical factor with the H.E.S.S. solid angle is $\overline{J}=1.33\times 10^{23}$~GeV$^2$cm$^{-5}$. The resulting exclusion limits are plotted (pink dashed line).

Below $\sim 1$~TeV, the \Fermi\ results provide stronger limits than the H.E.S.S. results. In comparison with the flux sensitivity (Section~\ref{parag:indepedant}), the $\left\langle \sigma v \right\rangle$ upper limits also take into account the predicted integrated number of gamma rays in the instrument energy range. The predicted number of gamma rays per annihilation event in the \Fermi\ energy range is about $10^3$ times higher than the one in the H.E.S.S. energy range. This implies a \Fermi\ limit which is of the order of 10 times better than the one for H.E.S.S., despite the latter's stronger flux sensitivity. \Fermi\ and H.E.S.S. give complementary limits on $\left\langle \sigma v \right\rangle$ in the $10$~GeV - $100$~TeV mass range. Since the given limits on the cross section are a few orders of magnitude above the theoretical expectations for DM particles in the Minimal Supersymmetric extension of the Standard Model (MSSM), those expectations are not shown in the figures.

\begin{figure}[h!]
  \begin{center}
    \mbox{\hspace{0cm}\includegraphics[scale=0.5]{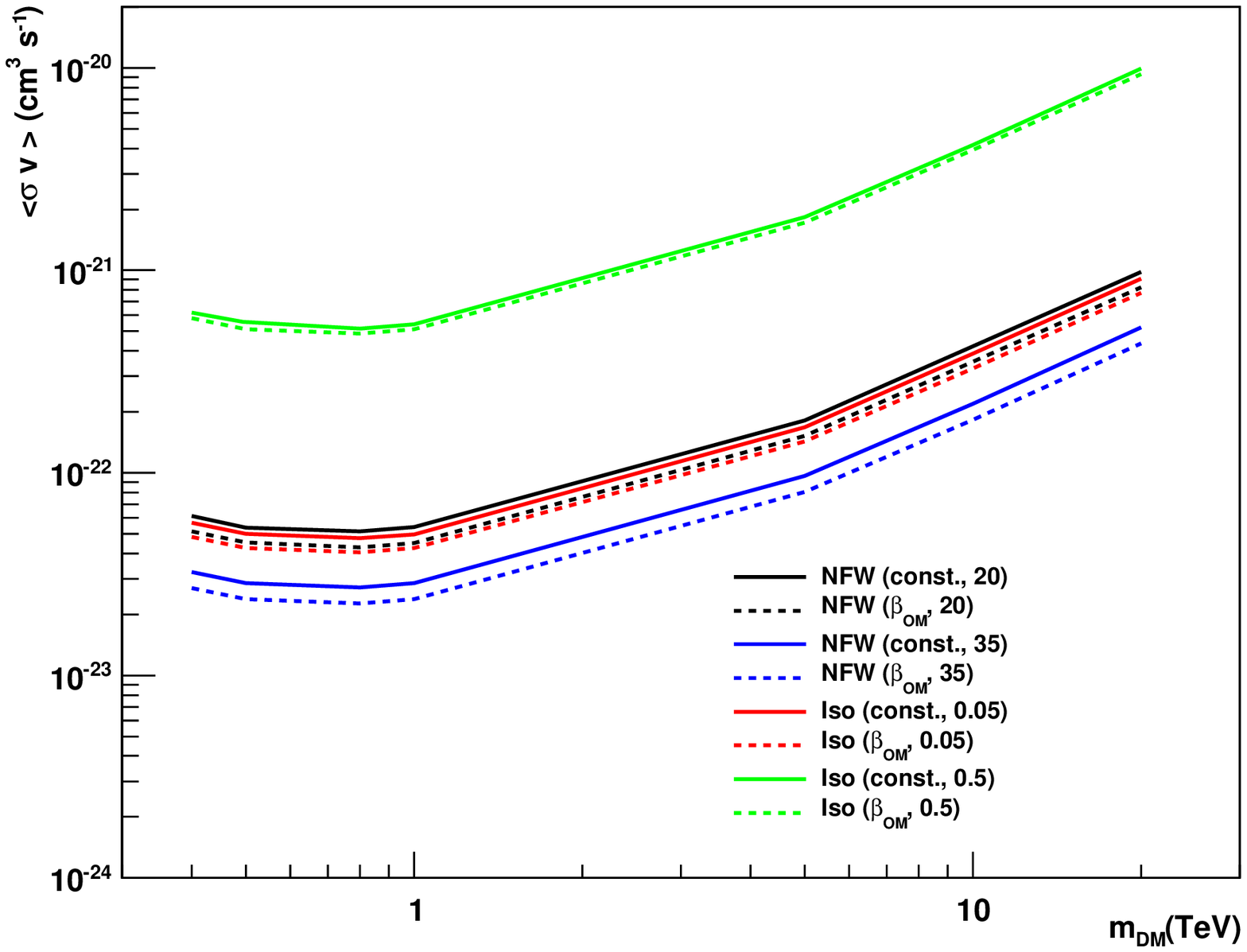}}
    \caption{Upper limit at $95\%$ C.L. of $\left\langle \sigma v \right\rangle$ as function of the Kaluza-Klein DM particle mass for all DM profiles of Sculptor in Table~\ref{tab:sculptor}. The velocity anisotropy and the concentration parameters are given in brackets for the NFW profile. The velocity anisotropy and the core radius are given in brackets for the pseudo-isothermal profile.\label{fig:exclusionKK}}
  \end{center}
\end{figure}

The Figure~\ref{fig:exclusionKK} shows the exclusion limits of $\left\langle \sigma v \right\rangle$ in the case of the Kaluza-Klein DM particle $\widetilde{B}^{(1)}$. The limits are plotted for the Sculptor dSph referring to the halo profiles given in the Table~\ref{tab:sculptor}. In the TeV range the $95\%$ C.L. upper limit on $\left\langle \sigma v \right\rangle$ reaches $10^{-23}$~cm$^{3}$s$^{-1}$.

\subsection{Enhancement effects for the exclusion limits \label{parag:boost}}

 Three cases that can modify the exclusion limits are considered: two particle physics effects, namely the \emph{Sommerfeld enhancement} and the \emph{Internal Brems\-strahlung} (IB) from the DM annihilation, and an astrophysical effect due to the mass distribution of \emph{dark-matter sub-halos}.

\subsubsection{The Sommerfeld enhancement}

Here two new assumptions were made for the Sculptor's and Carina's DM halo composition. The first assumption is to assume the DM particle  annihilates to a W boson, which is the case when the neutralino is a pure wino~\cite{2000NuPhB.570..455M}. The second assumption is to assume that the DM mean velocity inside the halo is the same as for the stars (the mean velocity dispersion of the stars is $\sigma_{\rm V}\sim 10.0$~km/s for Sculptor and $\sigma_{\rm V}\sim 7.5$~km/s for Carina). The later assumption is plausible due to the relaxation time scales in the dwarf galaxies.

\begin{figure}
  \centering
  \label{fig:exclusionSomSc}\includegraphics[scale=0.45]{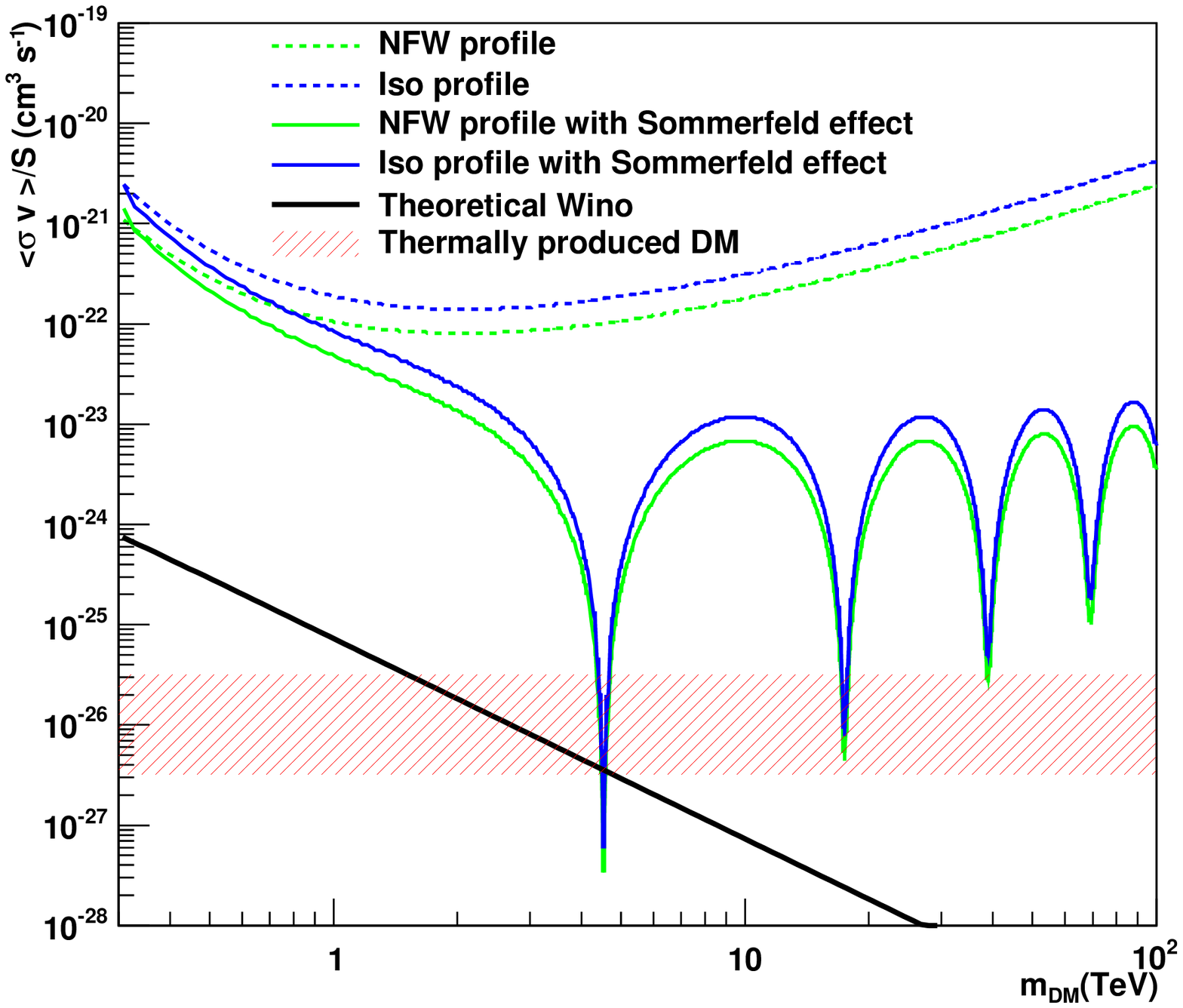}\\
  \label{fig:exclusionSomCar}\includegraphics[scale=0.45]{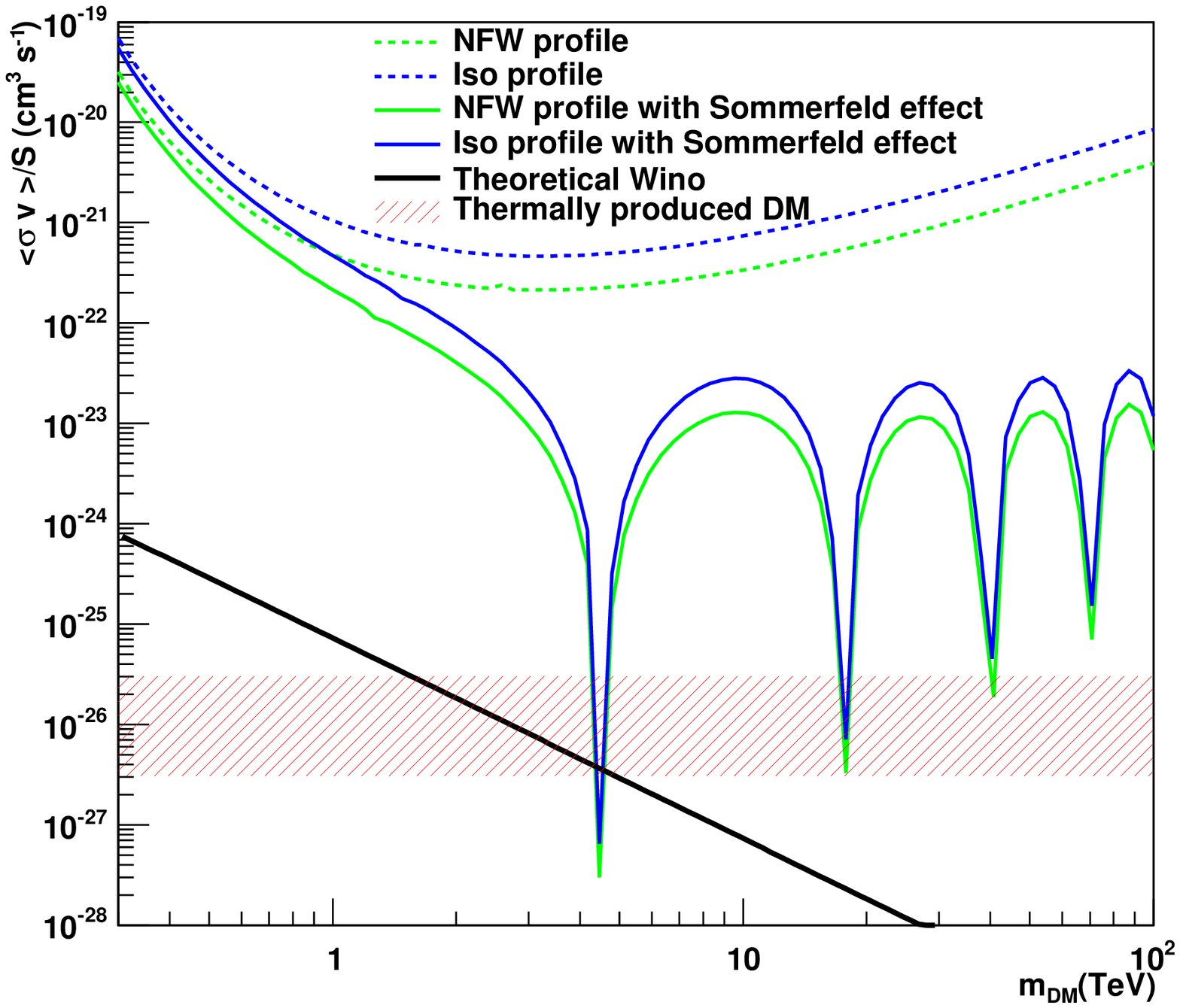}
  \caption{Upper limit at $95\%$ C.L. on $\left\langle \sigma v \right\rangle /S $ as function of the DM particle mass enhanced by the Sommerfeld effect (see text for more details) for Sculptor (\emph{top}) and Carina (\emph{bottom}). The NFW halo profile as well as the pseudo-isothermal profile are used. The predicted $\left\langle \sigma v \right\rangle_0$ for a pure wino~\cite{2000NuPhB.570..455M} (solid black line) as well as the typical cross section for a thermally produced DM (dashed red area) are also plotted.\label{fig:exclusionSom}}
\end{figure}

In this class of objects, the relative velocity between the DM particles may be sufficiently low so that the Sommerfeld effect can substantially boost the annihilation cross section~\cite{Lattanzi:2008qa}, since it is particulary effective in the very low-velocity regime. This non-relativistic effect arises when two DM particles interact in an attractive potential. The actual velocity-weighted annihilation cross section of the neutralino can then be enhanced by a factor S defined as
\begin{equation}
\left\langle \sigma v \right\rangle = S\left\langle \sigma v \right\rangle_0 \, ,
\end{equation}
where the value of S depends on the DM particle mass and relative velocity. A wino would interact with the attractive potential created by the Z gauge boson through the weak force before annihilation occurs, which would give rise to an enhancement. The value of this enhancement was numerically calculated as done in~\cite{Lattanzi:2008qa} and then used to improve the $95\%$~C.L. upper limit on the velocity-weighted annihilation cross section, $\left\langle \sigma v \right\rangle / S$ as a function of the DM particle mass. The effect of this enhancement is shown in Figure~\ref{fig:exclusionSom} for Sculptor (\emph{top}) and Carina (\emph{bottom}), and for two particular cases of the halo profile models. The predicted $\left\langle \sigma v \right\rangle_0$ for a pure wino~\cite{2000NuPhB.570..455M} as well as the typical annihilation cross section for a thermally produced DM ($\left\langle \sigma v \right\rangle_0$ $\sim 10^{-26}$~cm$^{3}$s$^{-1}$~\cite{1996PhR...267..195J}) are also plotted. Some specific wino masses can be excluded at the level of $\left\langle \sigma v \right\rangle_0$ $\sim 10^{-26}$~cm$^{3}$s$^{-1}$.

A more general case of the enhancement by the Sommerfeld effect can be treated by changing the exchanged boson mass and the coupling constant accordingly. The position of the resonances is qualitatively driven by equation~5 of~\cite{Lattanzi:2008qa}. Increasing the boson mass shifts the resonance to higher DM masses since the the weak coupling constant is only a slowly varying function of the boson mass. Simultaneously, once the relative velocity and the DM particle mass are fixed, the value of the enhancement close to the resonance grows roughly linearly with the boson mass~\cite{Lattanzi:2008qa}. Examples of the enhancement for 0.1 to 1.0  GeV exchange boson masses are shown in~\cite{Essig:2010em}.

\subsubsection{Internal Bremsstrahlung}

In some specific regions of the MSSM parameter space, the electromagnetic radiative correction to the main annihilation channels into charged particles can give a significant enhancement to the expected gamma ray flux in the observed environment due to internal Bremsstrahlung (IB)~\cite{Bergstrom:1989jr,Bringmann:2007nk}. In the \emph{stau co-annihilation region} of the minimal supergravity (mSUGRA) models, for instance, the wino annihilation spectrum would receive a considerable contribution from the internal Bremsstrahlung.

\begin{figure}[h!]
  \begin{center}
    \mbox{\hspace{0cm}\includegraphics[scale=0.45]{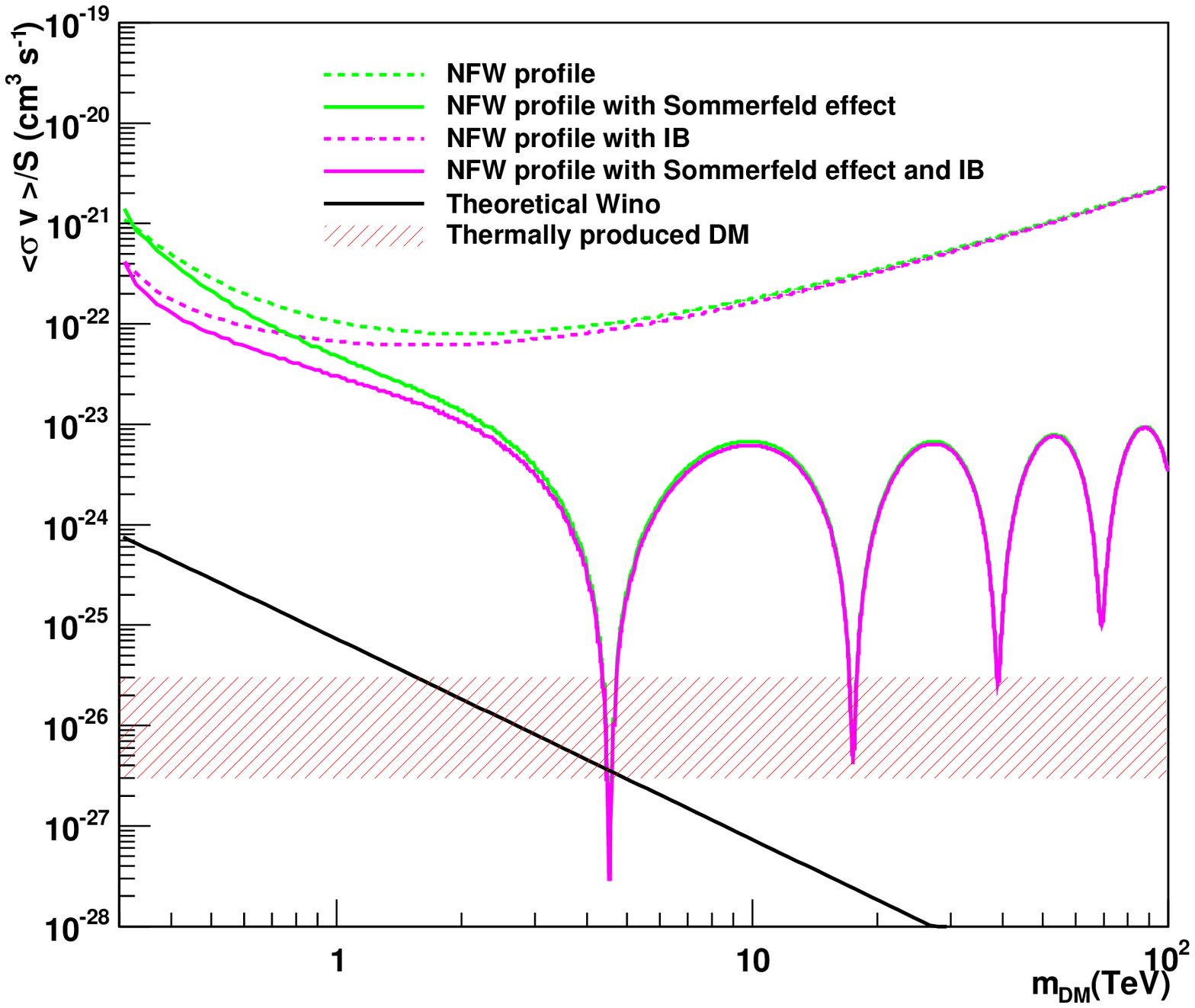}}
    \caption{Upper limit at $95\%$ C.L. on $\left\langle \sigma v \right\rangle /S $ as function of the DM particle mass enhanced by the Sommerfeld effect and the internal Bremsstrahlung  (see text for more details)  for a NFW profile of Sculptor. The predicted $\left\langle \sigma v \right\rangle_0$ for a pure wino (solid black line) as well as the typical cross section for a thermally produced DM (dashed red area) are also plotted.\label{fig:exclusionSom_IB}}
  \end{center}
\end{figure}

 This contribution to the annihilation spectrum was computed using the parametrization of~\cite{Bringmann:2007nk} for all the wino masses in the H.E.S.S. energy range. The enhancement effect on the $95\%$~C.L. upper limit on the velocity-weighted annihilation cross section is shown in Figure~\ref{fig:exclusionSom_IB}. The joint enhancement due to the Sommerfeld effect and IB is also plotted. The effect of the IB is only significant in the exclusion limits for the low mass DM particle regime.

\subsubsection{Enhancement from dark-matter sub-halos}

Astrophysical effects may also modify the exclusion limits. Numerical simulations of galactic halos predict a population of subhalos that could contribute to the overall astrophysical factor in equation~\ref{jbar}. Using the procedures given in~\cite{2009MNRAS.399.2033P}, the contribution to the astrophysical factor by the DM sub-halos population is estimated. An enhancement of the astrophysical factor is found to be of a few percent, which is too small to significantly affect the exclusion limits presented.

\section{Summary \label{sec:part6}}

Both Sculptor and Carina dSphs are well-studied in multiple wavelengths, providing reasonable measurements of the profile of the DM in their halos. Recent VHE gamma ray observations from H.E.S.S. of both of these objects provide new insight into the dark matter within them. While no positive dark-matter annihilation signal has been detected, the observations provide constraining limits on dark matter parameters.

Constraints have been obtained for the velocity weighted annihilation cross section $\left\langle \sigma v \right\rangle$ as a function of the mass for neutralino and KK DM particles. Concerning Sculptor dSph upper limits on $\left\langle \sigma v \right\rangle$ have been obtained in the range of $\sim$$10^{-21}$~cm$^3$s$^{-1}$ to $\sim$$10^{-22}$~cm$^3$s$^{-1}$ for neutralinos and $\sim$$10^{-21}$~cm$^3$s$^{-1}$ to $\sim$$10^{-23}$~cm$^3$s$^{-1}$ for KK particles. From the different profile parameters investigated, much better limits are obtained for a NFW profile with a strong concentration parameter $c=35$ when compared to the limits obtained for a isothermal profile with a large core radius of $r_{\rm c} = 0.5$~kpc. Also for the Carina Sph better limits for $\left\langle \sigma v \right\rangle$ of neutralinos have been obtained for a NFW profile.

The DM halo model induces systematic uncertainties in the exclusion limits: the value of the astrophysical factor can vary over one order of magnitude for a given halo profile in the case of Sculptor. The results presented show that the DM particle models that could satisfy WMAP constraints on the Cold Dark Matter relic density~\cite{2010arXiv1001.4538K} cannot be tested. H.E.S.S. limits are comparable to the limits reported by MAGIC~\cite{Albert:2007xg} and VERITAS~\cite{Acciari:2010pja}, but weaker than those obtained by \Fermi~\cite{2010ApJ...712..147A} in the GeV mass range. Nevertheless, they are complementary to the \Fermi\, limits in the TeV range. Searches through neutrinos provide competitive limits. In the WW channel IceCube constraints on $\left\langle \sigma v \right\rangle$ lie at the level of $\sim$$10^{-22}$~cm$^3$s$^{-1}$~\cite{2010arXiv1012.0184D}. Finally assuming a resonance effect in the Sommerfeld enhancement, some specific wino masses can be excluded, and the first experimental constraints have been obtained on the Sommerfeld effect using H.E.S.S. data and DM annihilation spectra.


\section*{Acknowledgements}

We would like to thank Mark Wilkinson for fruitful discussions concerning the modelling of Carina dSph. The support of the Namibian authorities and of the University of Namibia in
facilitating the construction and operation of H.E.S.S. is
gratefully acknowledged, as is the support by the German Ministry
for Education and Research (BMBF), the Max Planck Society, the
French Ministry for Research, the CNRS-IN2P3 and the Astroparticle
Interdisciplinary Programme of the CNRS, the U.K. Particle Physics
and Astronomy Research Council (PPARC), the IPNP of the Charles
University, the South African Department of Science and Technology
and National Research Foundation, and by the University of
Namibia. We appreciate the excellent work of the technical support
staff in Berlin, Durham, Hamburg, Heildelberg, Palaiseau, Paris,
Saclay, and in Namibia in the construction and operation of the
equipment.

\bibliographystyle{model1-num-names}
\bibliography{biblio}

\end{document}